\documentclass[final]{PoS}
\usepackage{shrthnds}
\usepackage{subfigure}
\usepackage{setspace}
\usepackage{multirow}
\newcommand{\cc}{{\mc C}}
\newcommand{\ccd}{{\mc C}^\dag}

\title{Production of a KK-graviton in association with a boson via gluon fusion at the LHC}

\ShortTitle{Production of KK-gravitons in association with a boson via gluon fusion at the LHC}

\author{\speaker{Subhadip Mitra}, Ambresh Shivaji and Pankaj Agrawal
       \\
       Institute of Physics, Bhubaneswar 751 005, India\\
       E-mail: \email{smitra@iopb.res.in}, \email{ambresh@iopb.res.in}, \email{agrawal@iopb.res.in}}

\abstract{We discuss the processes where a KK-graviton ($G$) of the extra-dimensional
 models is produced in association with a color singlet boson via gluon fusion at the LHC. In particular, we consider
the processes $gg \to h G$, $  \gamma G$, $  Z G$. These processes occur
at one-loop through box and triangle diagrams. The cross-section for the
process  $gg \to \gamma G$  vanishes at the one-loop level. It can be understood
by introducing the charge conjugation transformations of the KK-graviton.
The processes $gg \to h G$, $Z G$ can be observed at the LHC if the KK-graviton
and the Higgs boson exist with appropriate properties.}

\FullConference{10th International Symposium on Radiative Corrections (Applications of Quantum Field Theory to Phenomenology)\\
                September 26-30, 2011\\
                Mamallapuram, India}

\begin{document}
\section{Introduction}
Various extra-dimensional extensions of the Standard Model (SM) \cite{
ADD,RS,RS2,Antoniadis,Appelquist} have attracted a lot of interest in the recent literature.
In these models, the number of spacial dimensions is assumed to be more than $3$
with the extra dimensions being hidden (compactified). Apart from the compactification mechanism,
different models differ on the size as well as the number of the extra
dimensions. Although, depending on the model, the SM fields can either propagate in the bulk or live
on a boundary of the bulk, gravity can freely propagate through the extra-dimensions.
In the low energy 4-D picture gravity is treated as an effective theory with
the graviton fields  appearing as towers of KK-excitation modes (KK-gravitons).

The Large Hadron Collider (LHC) provides us with a unique opportunity to observe experimental signatures of these KK-gravitons.
For possible signals, several people have studied different production processes of spin-2 KK-gravitons (referred to as graviton in
the rest of the paper) in association with some vector boson in the LHC \cite{Kumar:2010kv,Kumar:2010ca,Karg:2009xk,Gao:2009pn}. Except for the case
where the final state vector boson is a gluon, these papers
consider only the $q\bar q$ initiated processes.
Here our focus is on a different initial state -- we discuss the two gluon initiated graviton production in association with a scalar/vector boson ($gg \rar G\mc B $).
Since gravitons couple with matter via energy-momentum tensor, only the $gg\rar Gg$ process has a tree level contribution.
For all the other bosons the corresponding process mediates via quark loops. We restrict
ourselves to color singlet final states and consider the following processes -- (i) $gg\rar Gh$, (ii) $gg\rar G\g$ and (iii) $gg\rar GZ$.

At the LHC, the gluon flux dominates over the quark-flux. Hence, although loop mediated, gluon fusion
contributions to the processes with the color singlet bosons need not be negligible. For the Higgs case, gluon fusion is the dominant channel
in the LHC. The cross-sections for this process in two different extra-dimensional models like ADD \cite{ADD} and RS1 \cite{RS}
have been reported earlier \cite{Shivaji}. In this paper we briefly describe the calculation and
summarize the results.
For the case of photon, however, gluon fusion gives zero contribution. This follows from the
introduction of C-parity of the graviton. We present a
small field theoretic proof of this argument. We present some results for the $Z$-boson case -- the details of this calculation
will be reported elsewhere \cite{ambresh}.

\section{Graviton with a Higgs Boson}

\begin{figure}[]
\bc
\subfigure[]{\includegraphics[width=0.195\textwidth]{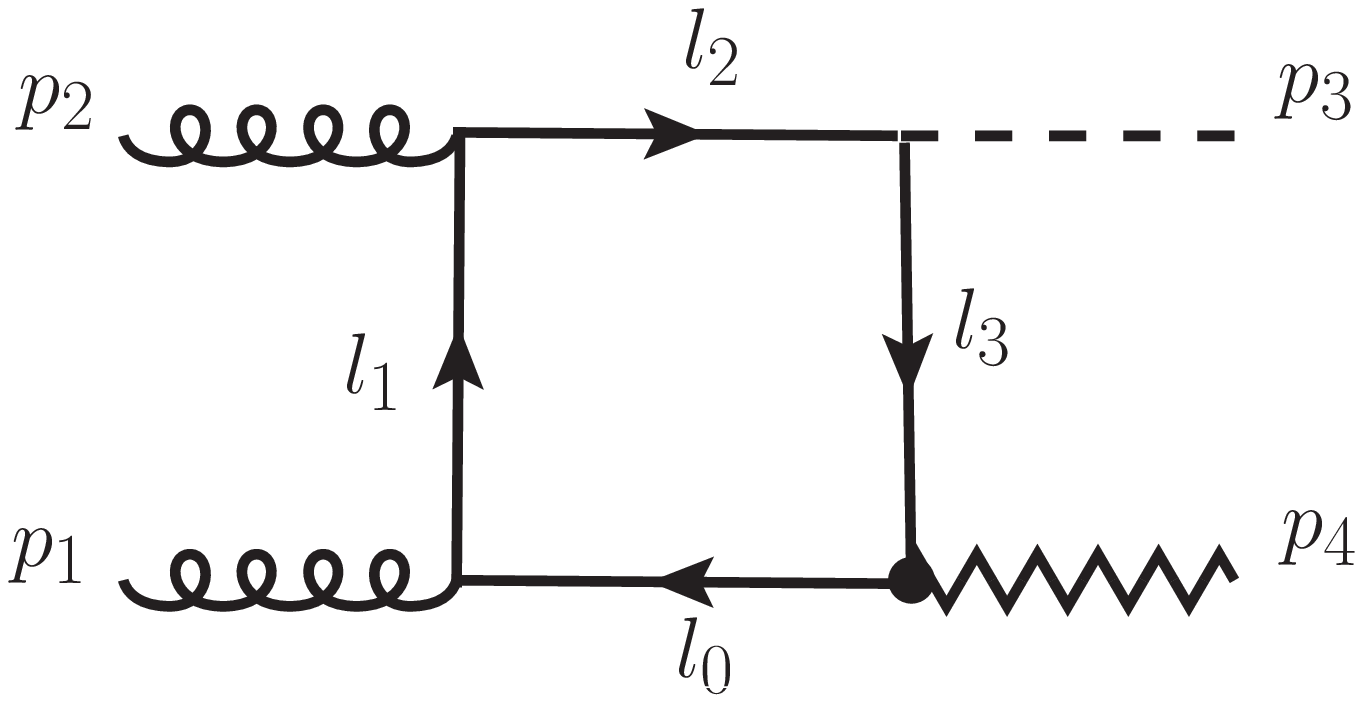}\label{fig:diaga}}
\subfigure[]{\includegraphics[width=0.195\textwidth]{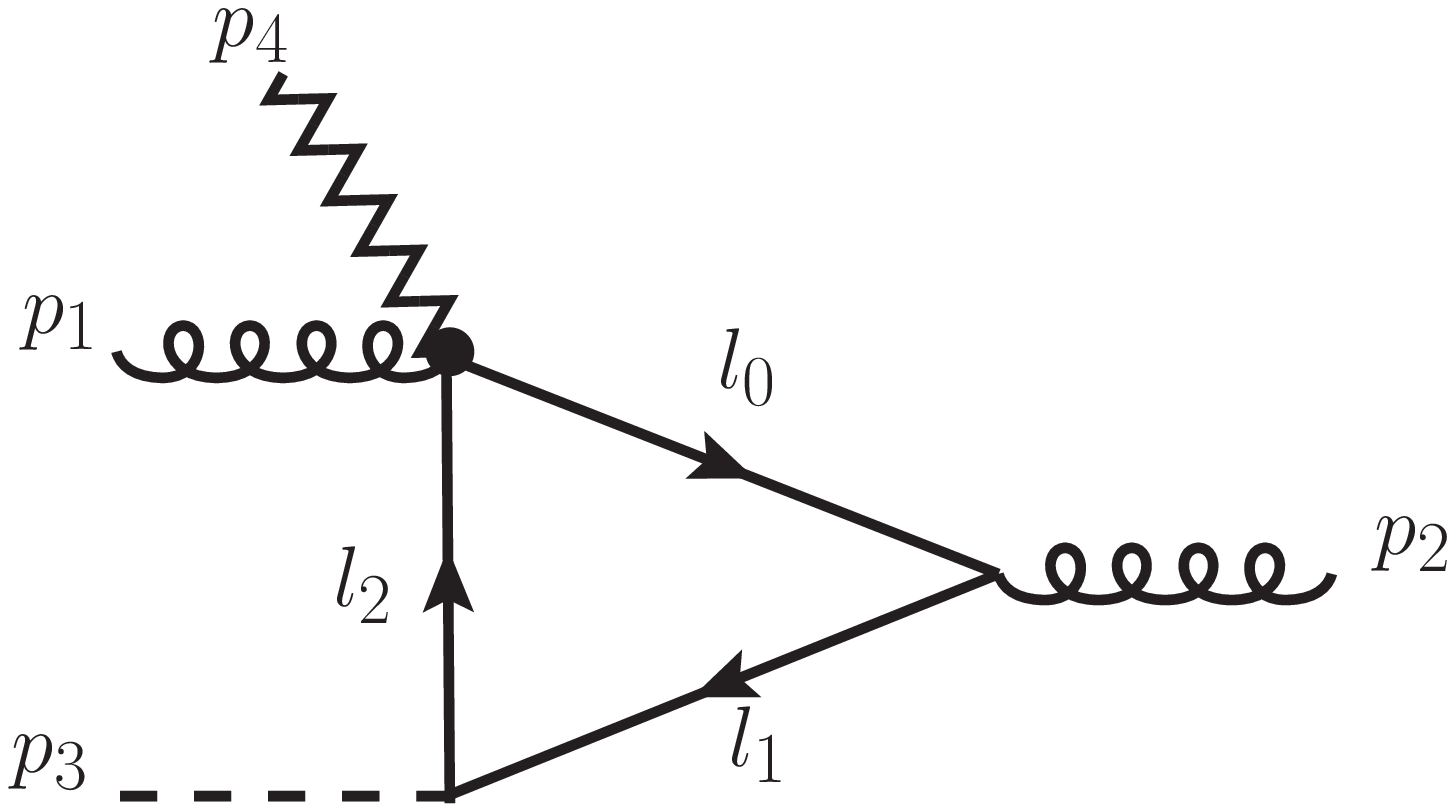}\label{fig:diagb}}
\subfigure[]{\includegraphics[width=0.195\textwidth]{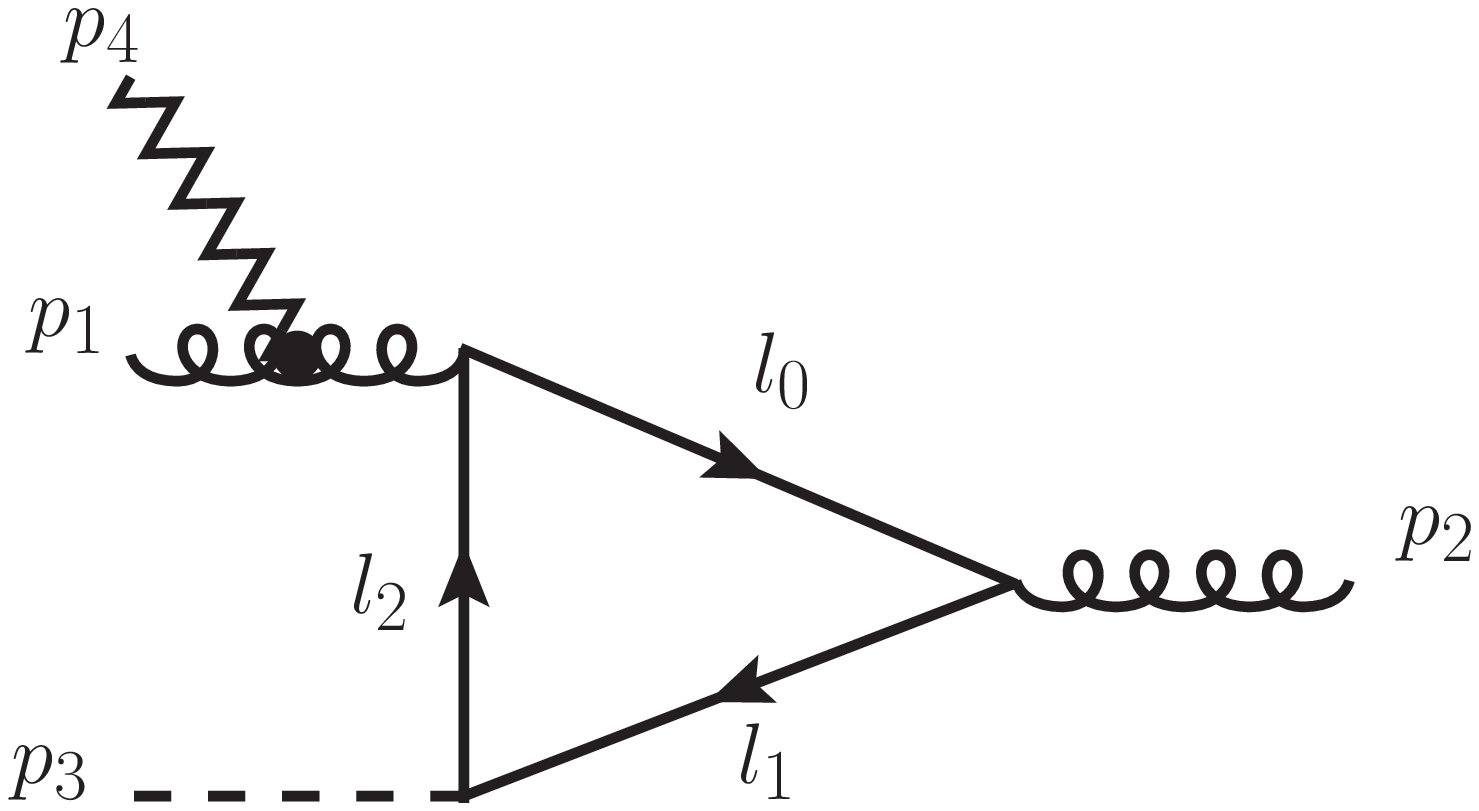}\label{fig:diagc}}
\subfigure[]{\includegraphics[width=0.195\textwidth]{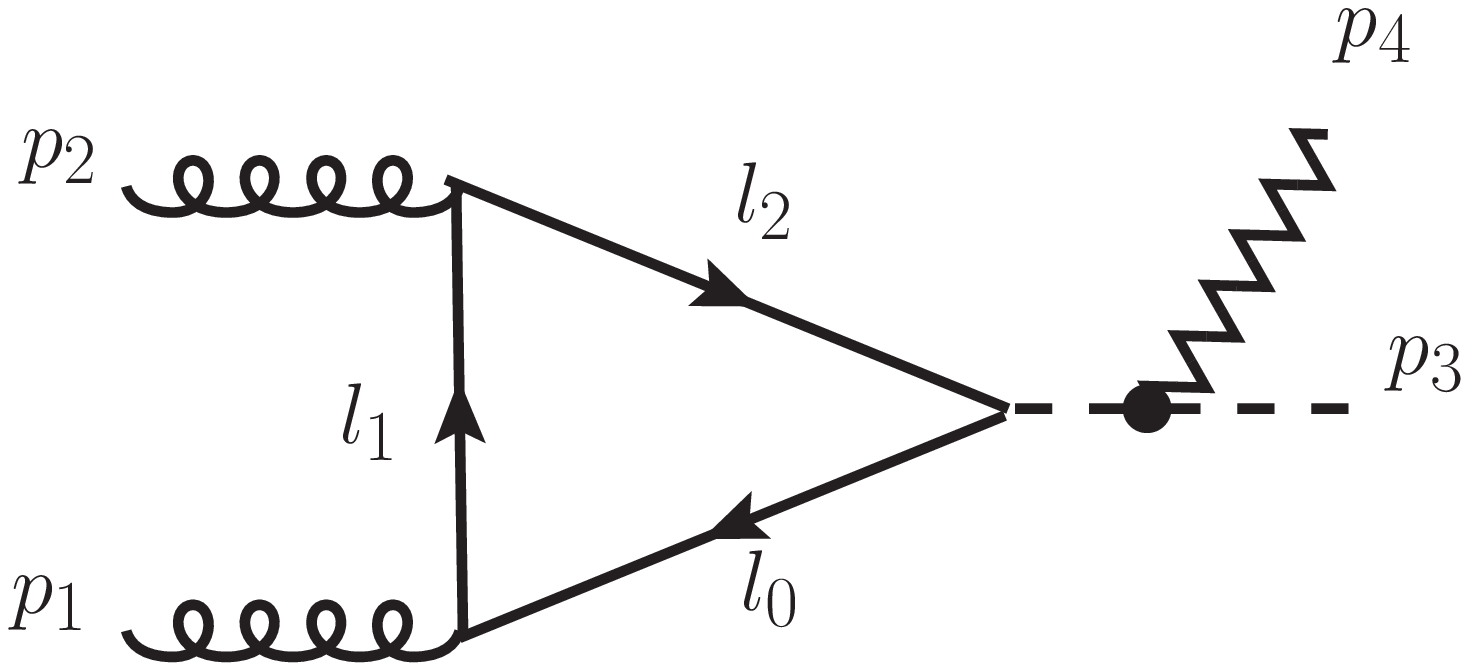}\label{fig:diagd}}
\subfigure[]{\includegraphics[width=0.195\textwidth]{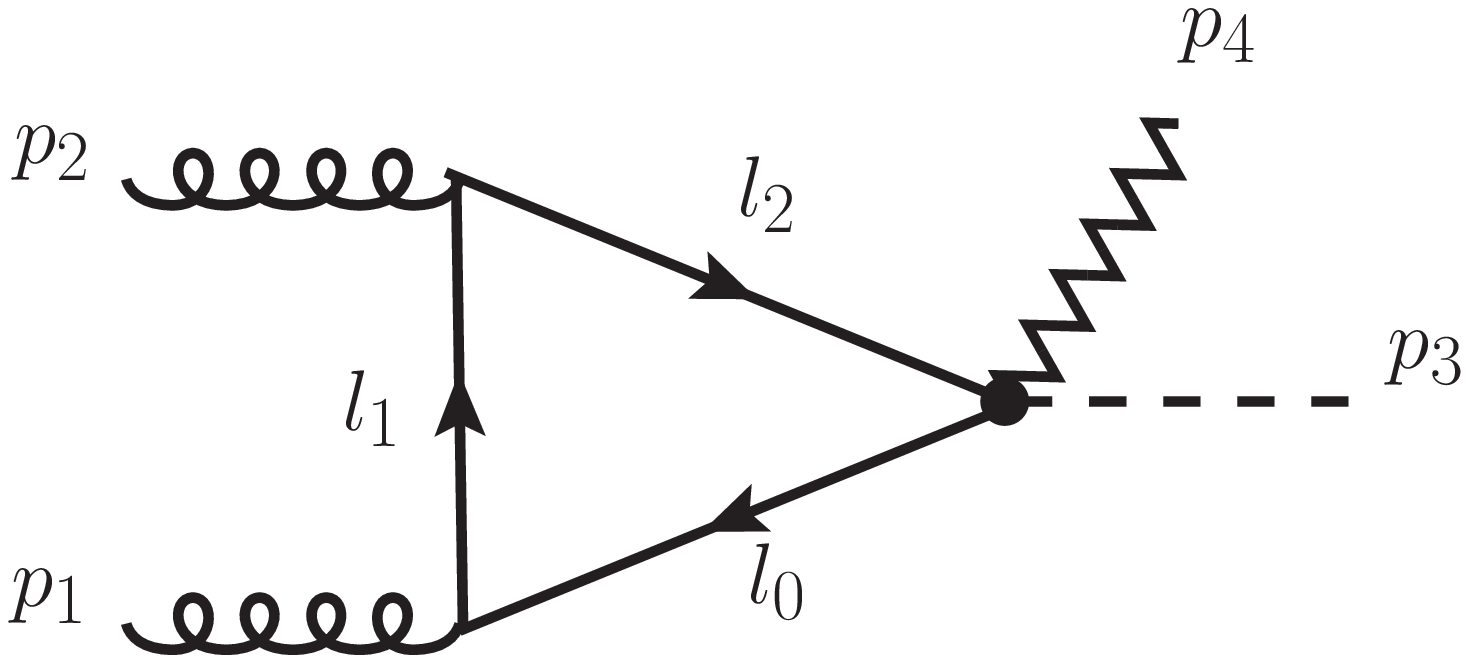}\label{fig:diage}}
\ec
\caption{Different classes of diagrams for the process $gg\rar h G$. The complete set of diagrams can be obtained by permuting the external particles.
The zigzag lines denote the graviton and the dashed lines denote the Higgs.}
\end{figure}

As already mentioned,  the gluon fusion mechanism ($gg\rar h G$)
is the dominant channel for the production of a Higgs boson in association with a graviton at the LHC \cite{Shivaji}.
Since both the final state particles are
color singlet, diagrams containing three gluon vertices are absent because of the color conservation.
The first non vanishing contribution to the $gg\rar h G$ process
comes from the diagrams containing a quark loop (at $\mc O(g^2_s \kp y_q)$). However,
 because of the presence of the Yukawa coupling ($y_q$) only the top-quark loop contributes significantly.
There are six box diagrams and twelve triangle diagrams (see Figs. \ref{fig:diaga} -- \ref{fig:diage}),  of which only half are independent
as they are related to the others by charge conjugation. Moreover the contribution from the triangle diagrams with a $hqqG$
vertex (Fig. \ref{fig:diage}) vanishes -- this vertex is proportional to the metric, $\et_{\m\n}$, which when contracted with the graviton polarization
tensor gives zero.

\subsection{Computation}
Feynman rules for the vertices required to calculate these diagrams can be found in \cite{Han}.
After computing the traces associated with the top-quark loops using  FORM \cite{Vermaseren:2000nd},
the amplitude contains tensor loop integrals, the most complicated of which are the
rank-4 tensor-box integral ($D^{\mu \nu \rho \sigma}$) among the the box integrals while rank-2 tensor-triangle integral ($C^{\mu \nu}$) among the triangle ones,
\begin{equation}
D^{\mu \nu \rho \sigma} = \int \frac{d^n l_0}{(2 \pi)^n} \frac{l_0^{\mu} l_0^{\nu} l_0^{\rho} l_0^{\sigma}}{D_0 D_1 D_2 D_3}\,,\quad
C^{\mu \nu} = \int \frac{d^n l_0}{(2 \pi)^n} \frac{l_0^{\mu} l_0^{\nu} }{D_0 D_1 D_2 }\,,
\end{equation}
where $D_i = l_i^2- m_t^2 + i\ve$ and $n=4-2\ep$ (see Figs. \ref{fig:diaga} - \ref{fig:diage} for the definition of $l_i$'s).
These tensor integrals were reduced into the standard scalar integrals -- $A_0$, $B_0$, $C_0$ and $D_0$
 using fortran routines \cite{agrawal} that follows the reduction scheme
developed by Oldenborgh and Vermaseren \cite{vanOldenborgh:1989wn}.
The scalar integrals (with massive internal lines) were ultimately called from FF library \cite{vanOldenborgh:1990yc}.
Helicity basis for the polarization vectors were used to calculate the amplitude.

To compute the cross-section, numerical integrations were performed over
the two body phase space, momentum fractions ($x_{1}/x_2$) of the initial state gluons and
over the graviton mass parameter in the continuum approximation (for the ADD model \cite{ADD,Han}).
As a cautionary check, the following tests were made with the code.
\begin{enumerate}
\item {\it UV Finiteness:}
The UV finiteness of the total amplitude were tested by varying the renormalization scale ($\m_{\rm R}$)
over ten orders of magnitude. The amplitude is independent of the actual value of $\m_{\rm R}$.
The triangle and box amplitudes are separately UV finite. Each triangle
diagram is UV finite by itself.

\item {\it Gauge Invariance:}
The amplitude was ensured to be gauge invariant with respect to both the gluons. This was done by replacing the polarization
vector of either of the gluons by its momentum ($\ve^\m(p_i) \rar p^\m_i$) which made the amplitude
vanish. Some of the triangle diagrams are separately gauge invariant with respect to both
the gluons. To ensure the correctness of their contribution towards the full amplitude,
gauge invariance check with respect to the graviton polarization was also performed.
\end{enumerate}

\subsection{Results}

\begin{figure}[]
\bc
\subfigure[]{\includegraphics [angle=0,width=.49\linewidth] {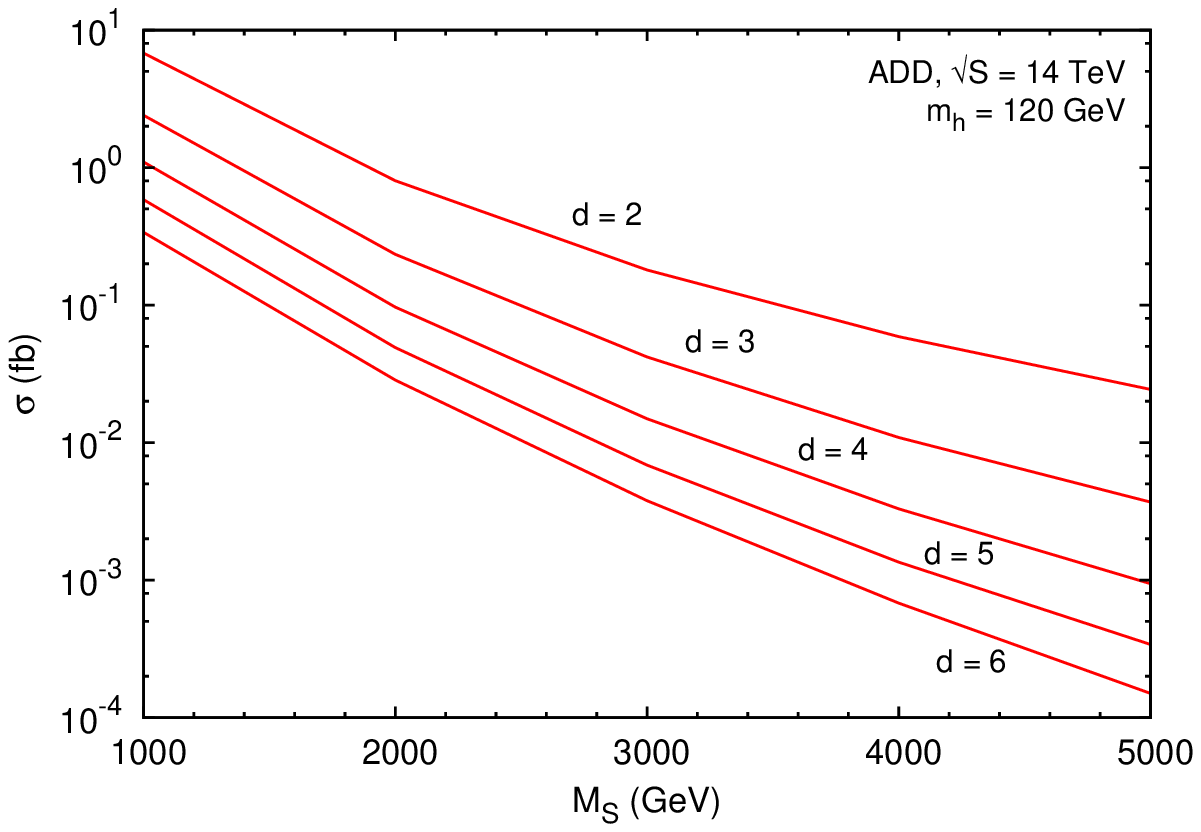}\label{fig:sigma_ms}}
\subfigure[]{\includegraphics [angle=0,width=.49\linewidth] {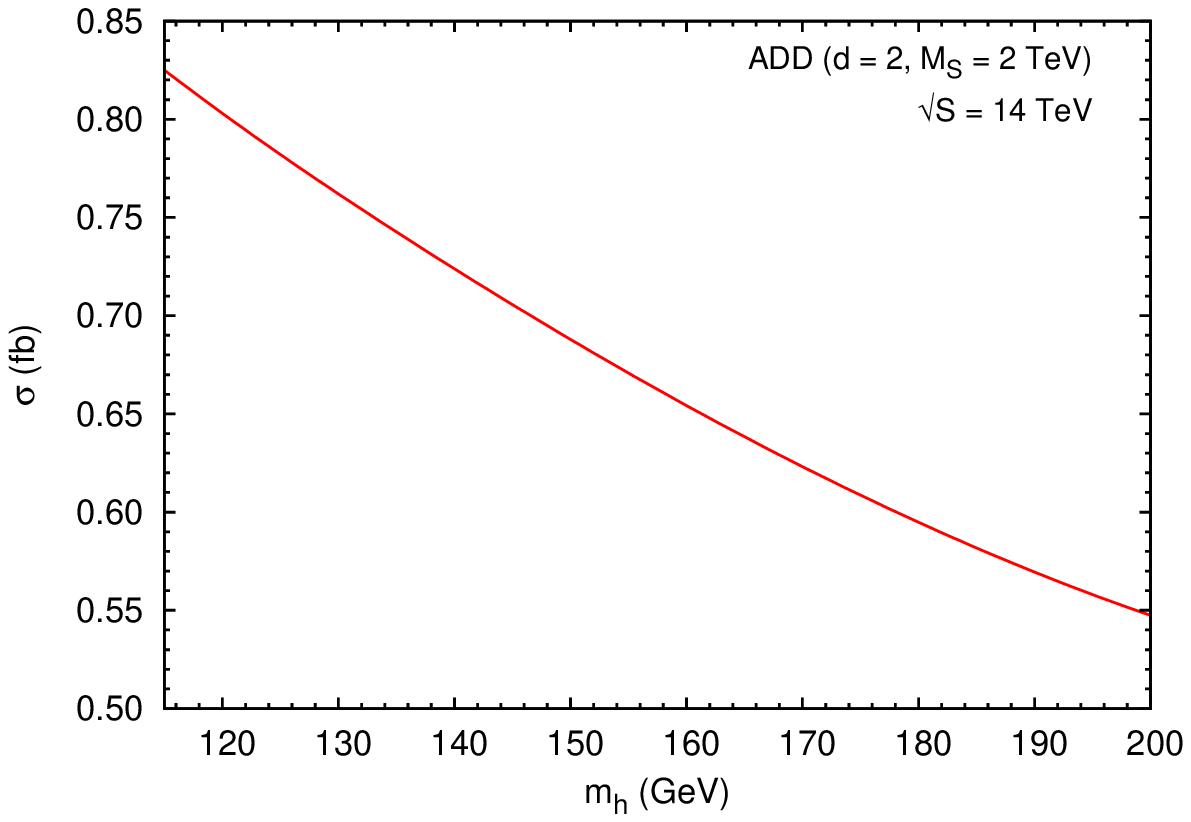}\label{fig:sigma_mh}}
\subfigure[]{\includegraphics [angle=0,width=.49\linewidth] {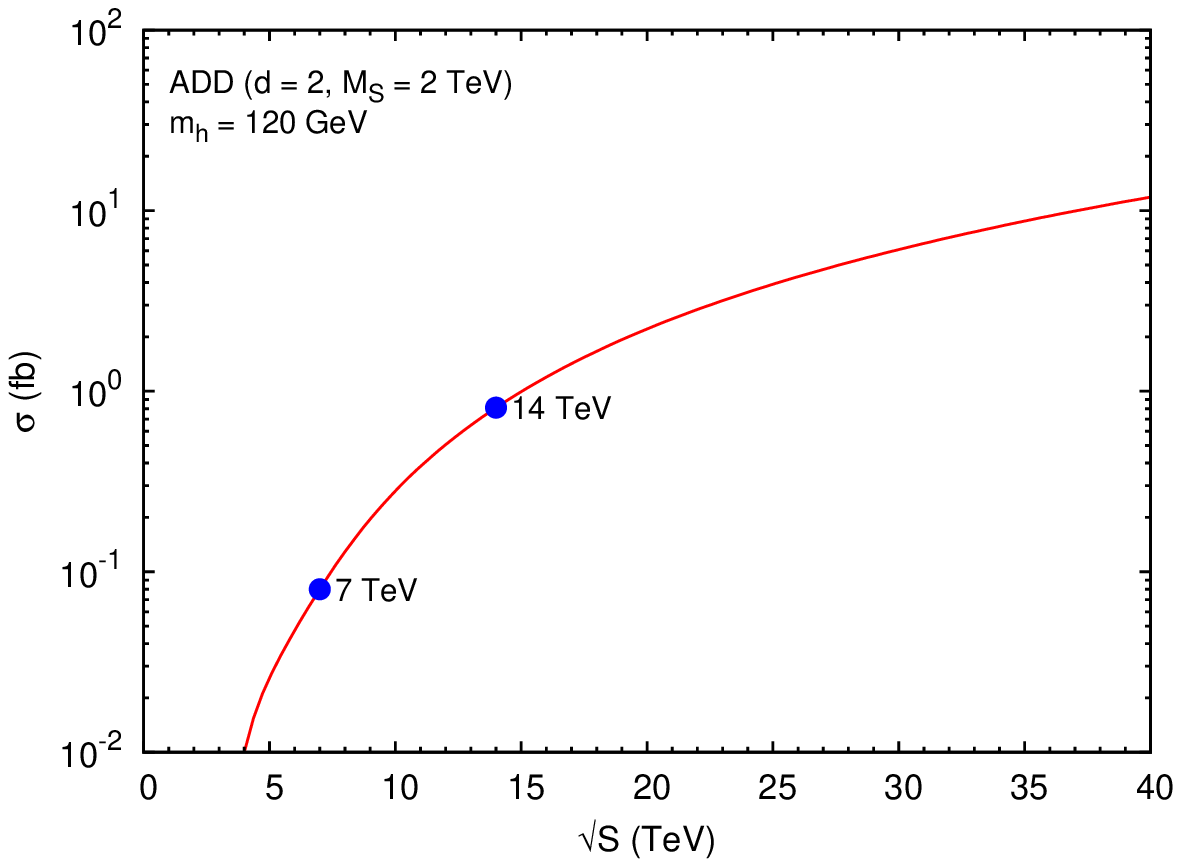}\label{fig:sigma_E}}
\subfigure[]{\includegraphics [angle=0,width=.49\linewidth] {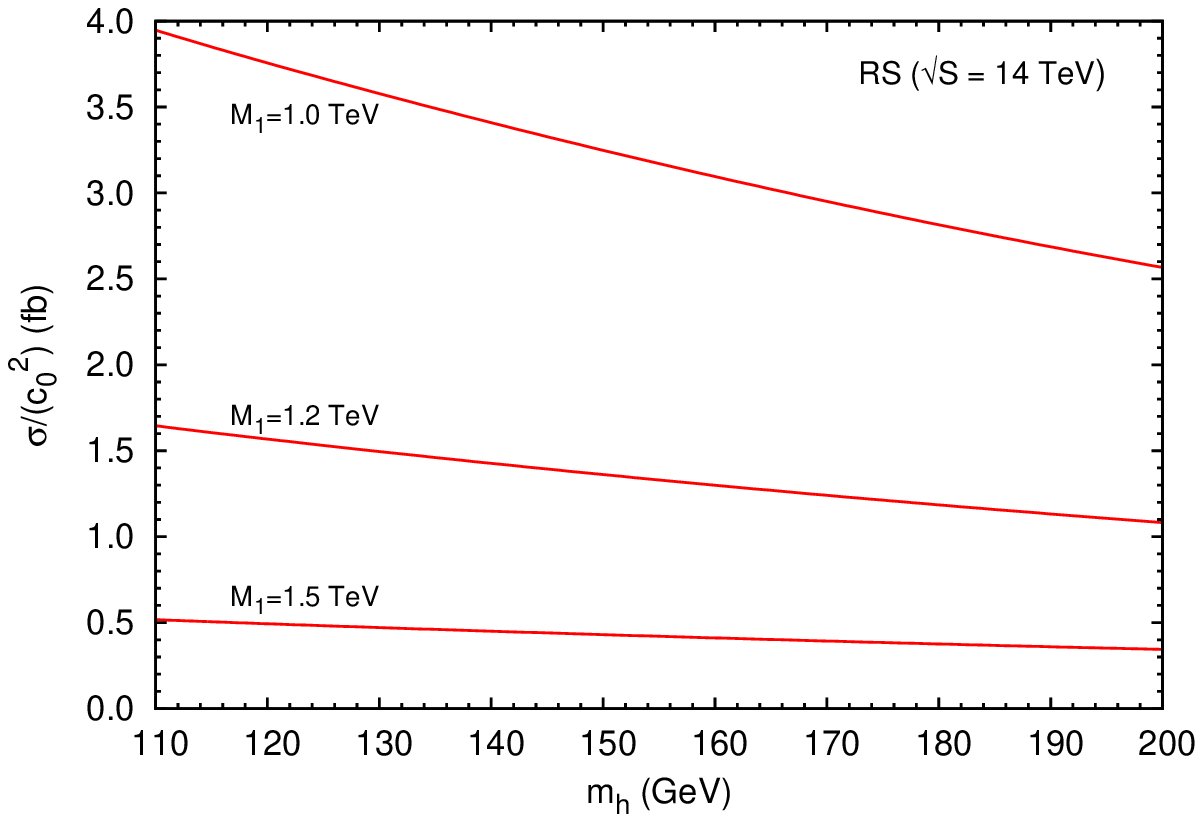}\label{fig:sigma_rs}}
\ec
\caption{Process $gg\to Gh$:  (a) Variation of the cross-section ($\s$) with the cutoff scale ($M_{\rm S}$) of the ADD model \cite{ADD} for different numbers of extra
dimensions, $d$. (b) Variation of $\s$ with mass of the Higgs, $m_h$. (c) Variation of $\s$ with $E_{cm}$.  (d) The cross-section, $\s$
for the production of the first KK mode of the graviton with the Higgs  scaled by the square of the dimensionless coupling
parameter $c_0= k/{\bar M}_{\rm P} = \sqrt{8\pi} k/M_{\rm P}$ in the RS1 model \cite{RS}. Here $M_1$ denotes the mass of the first KK mode of the graviton. }
\end{figure}

In Figs. \ref{fig:sigma_ms} -- \ref{fig:sigma_rs} we display the results for the KK-graviton production in association with a Higgs boson.
Fig. \ref{fig:sigma_ms} shows the dependence of the cross-section on the two parameters of the ADD model \cite{ADD}, {\it i.e.},
the cut-off scale, $M_{\rm S}$ and the number of extra dimensions, $d$. Fig. \ref{fig:sigma_mh} shows how the cross-section goes
down (mainly  because of phase space suppression)
with increasing Higgs mass and Fig. \ref{fig:sigma_E} shows the dependence of $\s$ on the center of mass energy, $E_{cm}$. In Fig. \ref{fig:sigma_rs}
we show the cross-section in the RS1  model  \cite{RS} scaled by the square of the dimensionless coupling parameter $c_0= k/{\bar M}_{\rm P} = \sqrt{8\pi} k/M_{\rm P}$.
In general $c_0$ is assumed to be between 0.01 and 0.1. These plots were obtained using NLO CTEQ6M PDFs and applying the following cuts on the transverse momentum and rapidity of the
Higgs: $p_{\rm T}^h > 20$ GeV, $|\eta^h| < 2.5$. In case of the ADD model, one extra cut was applied on the invariant mass of the
outgoing particles: $M(hG) \leq M_{\rm S}$ (truncated scheme).

We found a large cancelation between the box-diagrams contribution and the triangle-diagrams contribution
that reduces the amplitude by two-to-three orders of magnitude. This, in turn, reduces the cross-section
to the order of  1 fb or smaller for most of the parameter ranges of the ADD model. Still, one could expect few
hundred such events after the LHC achieves its design luminosity if $d \sim 2-3$ and $M_{\rm S} \sim 1-3$ TeV.
This process can be observed at LHC with a few years of operation. However in the RS1 model, the cross-section 
becomes even smaller. For example for $c_0 = 0.075$, $M_1=1$ TeV and $m_h =120$ GeV the
cross-section is only about 0.02 fb.  

Finally, before we move on to the next section, one comment on the effect of the mass of top quark is in order.
We find that the top quark does not decouple even as its mass, $m_t$ increases. In the beginning the cross-section increases  because of
the propagator enhancement. However, beyond $m_t \approx 400$ GeV, cross-section  
decreases and approaches a constant value beyond $m_t \gtrsim 2$ TeV. This behavior
is similar to what has been seen in the case of $gg \rar h$ production within the SM.

\section{Graviton with a Photon}

Photons do not have any charge (quantum number) and hence they are eigenstates of the Charge Conjugation (CC) operator $\mc C$.
Invariance of the QED Lagrangian under CC implies,
\ba
\cc A_{\m}\ccd = \et_\g A_{\m}\,,\quad \et_\g = -1\,,\label{eq:ccphoton}
\ea
{\it i.e.}, photons have negative C-parity. As a result there is no process with only odd number of external photons in QED. This is known as the
Furry's Theorem \cite{Furry, Nishijima}. To construct a field theoretic proof of this theorem let us consider the photon $n$-point Green's function,
\ba
\G_{\m_1\cdots\m_n} = \frac{1}{\mc N}  \left<0\left| {\sf T} \left[ A_{\m_1}(x_1)A_{\m_2}(x_2)\cdots A_{\m_n}(x_n)
\exp\left[i\int d^4x\, \mc L^{\rm QED}_{int}\right]\right]\right|0\right>\,,
\ea
where $\mc N$ is the normalization factor.
As $\ccd\cc = 1$,
\ba
\G_{\m_1\cdots\m_n} &=& \frac{1}{\mc N}  \left<0\left| {\sf T} \left[ \ccd\cc A_{\m_1}(x_1)\ccd\cc\cdots \ccd\cc A_{\m_n}(x_n)\ccd\cc
\exp\left[i\int d^4x\, \mc L^{\rm QED}_{int}\right]\ccd\cc\right]\right|0\right>\nn\\
&=& (-1)^n\G_{\m_1\cdots\m_n}\,,
\ea
where we have used Eq. \ref{eq:ccphoton} and the fact that both the free vacuum and QED interactions are invariant under CC.
Hence for odd $n$, the Green's function vanishes. This proof shows that this result is valid at all orders of perturbation theory as long as
the interaction terms remain invariant under CC.\footnote{Since weak interaction breaks CC invariance, odd number of photons can couple via $W$-boson loop. However
three photon vertex still remains zero by Yang's theorem \cite{Yang}.} Moreover insertion of any number of C-even boson fields would not affect the result.

\subsection{C-parity of Gravitons}

We introduce C-parity of gravitons to examine processes involving gravitons,
photons and gluons.
To determine the C-parity of the gravitons, let us consider the graviton-electron interaction \cite{Han},
\ba
2\kp^{-1} \lag_e^{\vec n} =
- G_{\m\n}\bar \psi_e i\g^\m\pr^\n\psi_e
          - \frac12 \bar \psi_e i\g^\m\left(\pr^\n  G_{\m\n}\right)\psi_e\,.\label{eq:gravint}
\ea
As gravity couples only to the energy-momentum tensor, it is natural to assume that the gravitational 
interaction with matter, in particular with electron, is invariant
under CC.  Using CC properties of the Dirac fields and gamma matrices,
we can determine that gravitons have positive C-parity,
\ba
\cc G_{\m\n}\ccd =   \et_G G_{\m\n}\,,\quad \et_G = 1\,.
\ea
Therefore as discussed above, any process with only odd number of external photons and any number 
of external gravitons vanishes to all orders of perturbation theory,
as long as we include only CC invariant interactions.

\subsection{Furry's Theorem with Gravitons, Photons and Two Gluons}
Gluons carry color charges and hence are not eigenstates of $\cc$. One cannot expect Furry's theorem to work for process with only
 external gluons and indeed three gluon vertex exists even at the tree level. However,
since QCD interactions are invariant under CC one can derive a transformation rule for gluons \cite{Tyutin},
\ba
\cc g^a_\m(x) \ccd &=& - \left[\Lambda\right]^{a b} g^b_\m(x)\,,
\ea
where $\Lambda$ is a diagonal matrix with $\Lambda^2 = 1$. It is defined as,
\ba
\left[\tau^a\right]^{\rm T} = \left[\Lambda\right]^{a b}\tau^b\,,
\ea
where $\tau^a$'s are the $SU(3)_c$ generators. The Green's function for a process with only $n$ 
number of external photons, $m$ number of external gravitons and two external gluons,
\ba
\G_{\{\m_i\} \{\alpha_j \beta_j\}\n_1\n_2} = \frac{\dl^{ab}}{\mc N}
\left<0\left| {\sf T} \left[ A_{\m_1}\cdots A_{\m_n}G_{\alpha_1 \beta_1} \cdots G_{\alpha_m \beta_m} 
g^a_{\n_1}g^b_{\n_2}\exp\left[i\int d^4x\, \mc L_{int}\right]\right]\right|0\right>\,,
\ea
where $\dl^{ab}$ appears  because of the conservation of color.
Just like before we can incert $\ccd\cc$'s to get
\ba
\G_{\{\m_i\} \{\alpha_j \beta_j\}\n_1\n_2} = 
(-1)^n\left(\Lambda\right)^2 \G_{\{\m_i\}\{\alpha_j \beta_j\}\n_1\n_2} = (-1)^n\G_{\{\m_i\} \{\alpha_j \beta_j\}\n_1\n_2}\,,
\ea
if $\mc L_{int}$ is invariant under charge conjugation. We see that the proof still works
if we replace any two photons by gluons, {\it i.e.}, force the two gluons to go into a color singlet state.
Hence two gluons can not fuse into odd number of photons (or any C-odd boson) and any number of 
gravitons or any other C-even boson. This is strictly true at one loop level. However as already 
mentioned this result is valid at all orders of perturbation theory as long as
we don't include weak corrections, {\it i.e.}, the interaction terms remain invariant under CC.

\section{Graviton with a $Z$-boson}
\begin{figure}[t]
\bc
\includegraphics[angle=0,width=.49\linewidth]{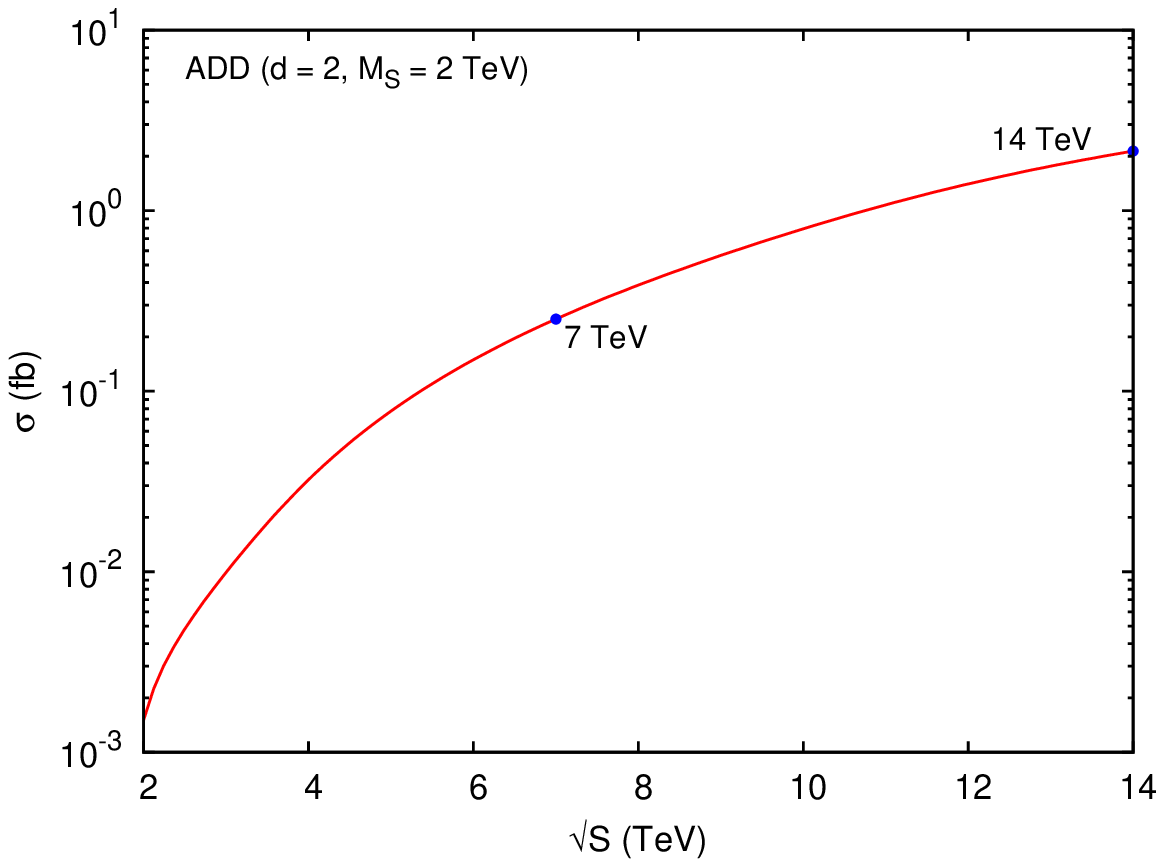}
\ec
\caption{Variation of $\s$ with $E_{cm}$ for the process $gg\to GZ$.}\label{fig:sigma_zg}
\end{figure}
     The production of a graviton in association with a Z-boson can occur
 at the tree level. One loop QCD corrections to this process have also been computed \cite{Kumar:2010kv,Kumar:2010ca}.
 However, this radiative correction calculation did not include the the gluon fusion channel. 
 Because of the large gluon flux at the LHC, this production can make sizable correction
 to the tree-level contribution. This calculation has recently been performed \cite{ambresh}.
 
The diagrams that contribute to the process $gg \to Z G$ belong to the same classes of the triangle and box diagrams as for
the process $gg \to h G$. Because of the Furry's theorem only the axial coupling of the 
Z-boson contributes to the amplitude in this process.
There is also an additional complication -- although this process is UV finite, the triangle and box diagrams are
 linearly divergent. Moreover the coupling to axial gauge current leads to
 anomalies. Therefore, one has to be careful in carrying out the computation and 
 checking the gauge invariance. The contribution of the individual flavor of quarks
 will give anomalous contribution, but this contribution must cancel when we
 include the full generation of quarks. One also has to treat $\gamma_5$ in $d$-dimensions
 more carefully and use proper prescription.

 The computation was done for the ADD model. The details of the calculation
can be found in \cite{ambresh}. In Fig. \ref{fig:sigma_zg}, we have plotted the cross-section of the process 
as a function of the center of mass energy for  $d = 2$ and $M_{\rm S} = 2$ TeV. For this,
the following kinematic cuts were applied:
$$
P_T^Z > 30 \textrm{ GeV},\quad |\eta^Z| < 2.5 ,\quad M(GZ) \leq M_{\rm S} \textrm{ (truncated scheme)}.
$$
The factorization and the renormalization scales were chosen as
$\mu_f = \mu_R = E_T^Z \left(=\sqrt{M_Z^2 + (P_T^Z)^2}\right)$ and, just like the Higgs case, NLO
 CTEQ6M PDFs were used. We note that at typical LHC
energy, the cross-section is of the order of few fb which is  much smaller than expected. 
The cross-section becomes small because of a two-orders of magnitude cancellation in the amplitude  between the box-type
and triangle-type diagrams. It is similar to the case of $gg \to h G$
process. Still one may expect few hundred of such events
after a few years of LHC operation at 14 TeV CM energy. 
Unlike the Higgs boson case where we don't find any decoupling of the heavy quark, 
the heavy quark in the loop does decouple as its mass goes to infinity for this process.

\section{Conclusions}

    We have examined the processes  $gg \to h G,  \gamma G,  Z G$ at the
  LHC. These processes, though leading order, occur at one loop. We have
  generalized the Furry's theorem to processes containing arbitrary number of photons,
  gravitons, and up to two gluons. According to this generalization, any process with
  only two gluons, odd number of photons and any number of gravitons vanish at
  one-loop order. This remains true to any order if we don't include CC-violating
  interactions, such as weak interaction. As a consequence, the process  
  $gg \to \gamma G$ does not
  get contribution at the one-loop level. In the calculation for the processes  
  $gg \to h G,  Z G$, there is a cancellation of  two orders of magnitude
  between the box and the triangle-classes of diagrams. This reduces the cross-sections
  to the order of 1 fb for these processes. Still, with few years of the operation
  of LHC at the center of mass energy of $14$ TeV, one may be able to observe these
  processes.
  
\section*{Acknowledgment}
We thank the organizers of RADCOR 2011 for their kind hospitality.

\end{document}